\documentclass{ws-procs9x6-cpt16}

\newcommand{\hf}{\frac{1}{2}}
\newcommand{\bea}{\begin{eqnarray}}
\newcommand{\eea}{\end{eqnarray}}
\newcommand{\nn}{\nonumber\\}

\providecommand{\Journal}[4] {#1 {\bf#2}, #4 (#3)}
\providecommand{\PRD}{Phys.\ Rev.\ D}
\providecommand{\ZP}{Z.\ Phys.\ }
\providecommand{\IJMPB}{Int.\ J.\ Mod.\ Phys.\ B}

\begin{document}

\newcommand{\refeq}[1]{(\ref{#1})}
\def\etal {{\it et al.}}

\title{The Impact of Lorentz Violation on the Klein Tunneling Effect}

\author{Zhi Xiao}

\address{Indiana University Center for Spacetime Symmetries\\
Bloomington, IN 47405, USA\\}

\address{Department of Mathematics and Physics,
North China Electric Power University\\
 Beijing 102206, China}

\begin{abstract}
We discuss the impact of a tiny Lorentz-violating $b^\mu$ term on the one dimensional motion of a Dirac particle scattering on a rectangular barrier.
We assume the experiment is performed in a particular inertial frame, where the components of $b^\mu$ are assumed constants.
The results show that Lorentz-violation modification to the transmission rate depends on the observer Lorentz nature of $b^\mu$. For a spacelike or lightlike $b^\mu$ the induced resonant frequency shift depends on the polarization, while for timelike $b^\mu$ there is essentially no modification.
\end{abstract}

\bodymatter

\section{Basic assumptions}
In this report, we try to use the Standard-Model Extension\cite{LVESM} to explore the impact of a tiny Lorentz symmetry violation (LV) $b^\mu$ on the Klein tunneling effect.\cite{Klein} The complete analysis is presented in Ref.\ \refcite{ZX}. For simplicity, we assume that we are working in a particular inertial reference frame, where all the components of $b^\mu$ remain constant. The hamiltonian of the LV modified Dirac equation is
\begin{equation}\label{FullbHam}
\hat{H}_b=\vec{\alpha}\cdot\hat{\vec{P}}+\gamma^0m-b^0\gamma^5+\vec{b}\cdot\vec{\Sigma}+U(Z),
\end{equation}
where $U(Z)=V_0[\Theta(Z)-\Theta(Z-L)]$ and $\Sigma^i\equiv\hf\epsilon_{ijk}\sigma^{jk}$, with $\epsilon_{ijk}$ being a totally antisymmetric tensor with $\epsilon_{123}=1$.
Next, we will use the spacelike $b^\mu$ term as an example to show the LV impact on the Klein tunneling, which lets us get a glimpse of how the LV impact depends on the observer Lorentz nature of $b^\mu$.

\section{The impact of the LV $b^\mu$ term on Dirac tunneling}\label{Main}
\begin{figure}
 \begin{center}
  \scalebox{1.0}{\includegraphics[width=0.7\hsize]{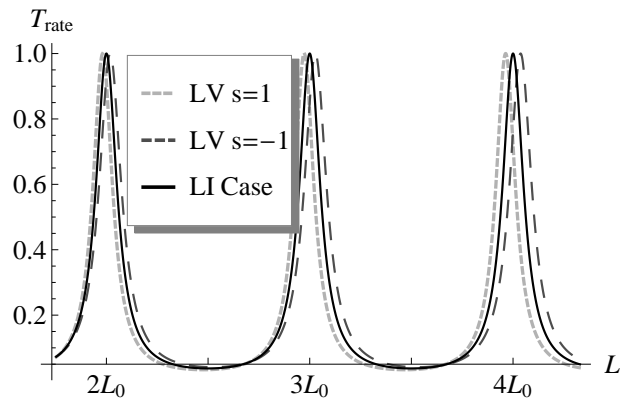}}
  \caption{Klein tunneling rate as a function of barrier length. The solid black curve corresponds to the Lorentz-invariant (LI) tunneling rate, while the dashed gray and black curves correspond to the Lorentz-violating tunneling rate with helicity $s=+1,-1$ respectively. The length unit is $L_0=\pi/\sqrt{(E-V_0)2-m^2}$.}\label{KTRL}
 \end{center}
\end{figure}

For spacelike $b^\mu$, we can use observer Lorentz symmetry to transform it into $b^\mu=(0,\vec{b})$, and further assume $\vec{b}(\theta,\phi)=b(\sin\theta\cos\phi,\sin\theta\sin\phi,\cos\theta)$ for convenience. The static wave function can take the form
\bea
\Psi_{R} (Z,T)
&=&e^{-iET}\left\{[\phi_i(p)e^{ipZ}+\phi_r(p')e^{-ip'Z}]\Theta(-Z)\right.\nn
&&
\hskip 30pt
\left.+[\phi_f(q)e^{iqZ}+\phi_g(q')e^{-iq'Z}][\Theta(Z)-\Theta(Z-L)]\right.
\nn
&&
\hskip 30pt
\left.+\phi_t(p)e^{ipZ}\Theta(Z-L)\right\},
\label{WaveF}
\eea
where $\phi_{n}=\left
(\begin{array}{c}
    \xi_{n} \\
    \eta_{n}
  \end{array}
\right)$ with subscript $n=i,r,f,g,t$ respectively. According to the Dirac equation $i\dot{\Psi}_R(Z,T)=\hat{H}_b\Psi_{R}(Z,T)$, each $\phi_n$ must satisfy
\bea\label{GSDirac}&&
\left(
  \begin{array}{cc}
    m-[E-U(Z)]+\vec{b}\cdot\vec{\sigma} & \vec{\sigma}\cdot\hat{\vec{P}} \\
    \vec{\sigma}\cdot\hat{\vec{P}} & [U(Z)-E]-m+\vec{b}\cdot\vec{\sigma} \\
  \end{array}
\right)\phi_n=0.
\eea
From the lower equation of (\ref{GSDirac}), we determine $\eta_{n}$ and
substitute it back into the upper equation of (\ref{GSDirac}) to get the dispersion relation
\bea\label{GSpaDR}
(p^2+m^2-E^2)^2-2b^2(p^2\cos2\theta+E^2+m^2)+b^4=0.
\eea
A similar relation applies to $q$ with the replacements $p\rightarrow{q},~E\rightarrow{E-V_0}$ in
Eq.\ \eqref{GSpaDR}.
Assume further that
$\xi_{n}\equiv\mathcal{N}\xi_i$, where
$\mathcal{N}=\mathcal{R},\mathcal{F},\mathcal{G},\mathcal{T}$ respectively. By
using the continuity equation at points $Z=0,~L$, in principle we can solve for all the proportionality constants $\mathcal{R},\mathcal{F},\mathcal{G},\mathcal{T}$.
For example, using the method above, we can determine the transmission amplitude as
\bea\label{Tramp}&&
\mathcal{T}=\frac{e^{-ipL}}{\cos[qL]-\frac{i}{2}(\frac{\mathcal{K}p}{q}+\frac{q}{\mathcal{K}p})\sin[qL]},
\eea
where $\mathcal{K}\equiv{1-\frac{V_0}{E+m-bs}},~s=\pm1$, and we have used $p=p',q=q'$ since
Eq.~\eqref{GSpaDR}
is an even function of $p$. Next, supposing $\vec{b}=b~\hat{e}_Z$, we get from
Eq.~\eqref{Tramp}
the resonant energy
\bea\label{ResEez}
E_{\rm Res}(n,s)=V_0+bs+\sqrt{m^2+(n\pi/L)^2}
\eea
when $E>V_0+m+bs$. This is the relativistic counterpart of ordinary quantum-mechanical resonant transmission.\cite{ZXMJPB} For a specific resonance number $n$, the resonant energy difference between states with opposite helicity is $\delta{E}_{\rm Res}(n)\equiv{E_{\rm Res}(n,+1)-E_{\rm Res}(n,-1)}=2b$.
For a specific barrier height $V_0$ and energy $E$ of the incoming electron, the resonant barrier length is $L(n,s)=\frac{n\pi}{\sqrt{(E-V_0-bs)^2-m^2}}$. The resonant-length difference between opposite-helicity states with the same $n$ is
    \bea&&\label{ResLeng}
    \delta{L}(n)=L(n,+1)-L(n,-1)\simeq2b\frac{E-V_0}{(E-V_0)^2-m^2}L_{\rm LI}(n),
    \eea
where $L_{\rm LI}(n)=\frac{n\pi}{\sqrt{(E-V_0)^2-m^2}}$ is the Lorentz-invariant resonant-barrier length. So the resonant-length difference $\delta{L}(n)$ increases with the barrier length $L$. In principle, we may use this and a precisely controllable long barrier to amplify the tiny LV $b^\mu$ effect. See Fig.\ \ref{KTRL}, where we plot with an impractically large $b\propto0.001\mathrm{m}_e$ to make the LV helicity-dependent shift of the resonant spectrum distinguishable.
A complete discussion using other types of $b^\mu$ is similar; see Ref.\ \refcite{ZX}.

\section*{Acknowledgments}

The author appreciates valuable discussions
with Alan Kosteleck\'y, Hai Huang, Ralf Lehnert, and Herb Fertig,
and is grateful for IUCSS hospitality
and for study-abroad funding from the Chinese Scholarship Council.


\begin{thebibliography}{x}

\bibitem{LVESM}
D.\ Colladay and V.A.\ Kosteleck\'y,
\Journal{\PRD}{55}{1997}{6760};
\Journal{\PRD}{58}{1998}{116002};
V.A.\ Kosteleck\'y,
\Journal{\PRD}{69}{2004}{105009}.

\bibitem{Klein}
O.\ Klein,
\Journal{\ZP}{53}{1929}{157}.

\bibitem{ZX}
Z.\ Xiao,
\Journal{\PRD}{93}{2016}{125022}.

\bibitem{ZXMJPB}
Z.\ Xiao, H.\ Huang, and X.X.\ Lu,
\Journal{\IJMPB}{29}{2015}{1550052}.

\end{thebibliography}
\end{document}